\documentclass[conference,compsoc]{IEEEtran}
\IEEEoverridecommandlockouts
\usepackage{cite}
\usepackage{amsmath,amssymb,amsfonts}
\usepackage{algorithm,algpseudocode}
\usepackage{graphicx}
\graphicspath{ {Figures/} }
\usepackage{textcomp}
\usepackage{xcolor}
\usepackage{hyperref}
\usepackage{lipsum}
\usepackage{tabularx}
\usepackage{booktabs}
\usepackage{nicefrac}

\def\BibTeX{{\rm B\kern-.05em{\sc i\kern-.025em b}\kern-.08em
    T\kern-.1667em\lower.7ex\hbox{E}\kern-.125emX}}
\begin{document}

\setlength\extrarowheight{2pt}

\title{Decentralised Trustworthy Collaborative Intrusion Detection System for IoT}

\author{
    \IEEEauthorblockN{
        Guntur Dharma Putra\IEEEauthorrefmark{1}\IEEEauthorrefmark{4},
        Volkan Dedeoglu\IEEEauthorrefmark{2},
        Abhinav Pathak\IEEEauthorrefmark{1}\IEEEauthorrefmark{5},
        Salil S. Kanhere\IEEEauthorrefmark{1}\IEEEauthorrefmark{4}
        and Raja Jurdak\IEEEauthorrefmark{3}
    }
    \IEEEauthorblockA{
        \IEEEauthorrefmark{1}UNSW, Sydney
        \IEEEauthorrefmark{2}CSIRO Data61, Brisbane
        \IEEEauthorrefmark{3}QUT, Brisbane
        \IEEEauthorrefmark{4}CSCRC, Australia
        \IEEEauthorrefmark{5}BITS, Pilani
    }
}

\maketitle

\begin{abstract}
Intrusion Detection Systems (IDS) have been the industry standard for securing IoT networks against known attacks. To increase the capability of an IDS, researchers proposed the concept of blockchain-based Collaborative-IDS (CIDS), wherein blockchain acts as a decentralised platform allowing collaboration between CIDS nodes to share intrusion related information, such as intrusion alarms and detection rules. However, proposals in blockchain-based CIDS overlook the importance of continuous evaluation of the trustworthiness of each node and generally work based on the assumption that the nodes are always honest. In this paper, we propose a decentralised CIDS that emphasises the importance of building trust between CIDS nodes. In our proposed solution, each CIDS node exchanges detection rules to help other nodes detect new types of intrusion. Our architecture offloads the trust computation to the blockchain and utilises a decentralised storage to host the shared trustworthy detection rules, ensuring scalability. Our implementation in a lab-scale testbed shows that the our solution is feasible and performs within the expected benchmarks of the Ethereum platform.
\end{abstract}

\begin{IEEEkeywords}
blockchain, IoT, intrusion detection system, collaborative, scalability, trust management
\end{IEEEkeywords}

\section{Introduction}
\label{sec:intro}
Intrusion Detection Systems (IDS) have been widely deployed as a means of securing IoT networks, with the goal of detecting malicious activity. In general, IDS can be categorised into two groups based on their underlying mechanics: signature-based; and anomaly-based. Signature-based IDS identifies incoming attacks by matching the network traffic with known intrusion signatures or rules. Anomaly-based techniques observe certain disparities in network traffic using machine learning approaches.

In recent years, an expansion of the attack surface has been inevitable, partially due to the adoption of IoT devices in diverse areas. This has consequently escalated the importance of defending the network from emerging threats~\cite{Anthi2019}. Unfortunately, conventional IDS that work in isolation may be easily compromised, since they are unaware of new attacks which are not in their detection database. Researchers have thus proposed utilising multiple IDS to work together, referred to as Collaborative-IDS (CIDS), in which each IDS node shares its expertise and experience, e.g., alarms and attack signatures, to build collective knowledge of the recent attacks and increase the detection accuracy~\cite{Fung2011}.

Researchers have recently explored the use of blockchain, the technology behind Bitcoin, to enhance the performance of CIDS. For instance, blockchain is utilised to provide a transparent layer for sharing detection signatures\cite{Li2019}. The decentralised nature of blockchain also ensures fault tolerance and removes the need of a fully trusted third party in managing collaboration between CIDS nodes~\cite{Liang2021}. A peer-to-peer consensus algorithm in block generation is employed to build a trusted database of CIDS detection models~\cite{Golomb2018}.

However, proposals of blockchain-based CIDS overlook the importance of continuous evaluation of each node's trustworthiness and generally work based on the assumption that the nodes are always honest. In fact, a trusted node may later be compromised and  share untruthful detection rules to contaminate the rule detection database which would potentially expose the network to attacks~\cite{9097612}. To achieve trustworthy and effective collaboration in CIDS, continuous evaluation of trustworthiness of CIDS nodes is required, which can be derived from their collaboration behaviour. The shared detection rules database would grow significantly as the collaboration continues, making scalability another key factor to consider. Additionally, there is a need for secure and transparent trust mechanisms with the goal of providing auditability.

In this paper, we propose a trustworthy CIDS architecture that continuously evaluates the trustworthiness of the CIDS nodes by evaluating the quality of the detection rules contributed by each IDS node. Each participating CIDS node can update its knowledge with the trustworthy detection rules to detect new attacks. We utilise a peer-to-peer decentralised storage to maintain a copy of the shared trustworthy detection rules, thus ensuring scalability. We divide the participating nodes into three categories, namely validator, contributor and regular nodes, each of which has a different role in the system. We also design two smart contracts, Trust and Reputation Management (TRM) and Storage smart contract, to quantify each node's trustworthiness and manage the decentralised storage. Our architecture offloads trust computation to the TRM smart contract which reduces the computation load for each CIDS node. We design a smart contract-based consensus algorithm to achieve collaborative detection rule validation and avoid an adversary from contributing deceptive detection rules. We design our architecture as a blockchain agnostic platform which can be implemented on any blockchain instantiation that supports smart contracts. While we use signature-based CIDS as an illustrative example, the trust management concept can be generalised to other types of CIDS.

In summary, we make the following contributions:
\begin{itemize}
    \item We propose a trustworthy CIDS architecture that continuously evaluates contributions from each CIDS node to protect the network from invalid detection rules. We present transparent and accountable trust mechanisms that provide auditability.
    \item Our proposed CIDS architecture offloads the trust computation and trustworthy detection rules to the blockchain and the decentralised storage, thus reducing the load on each CIDS node.
    \item Our trust mechanism separates the trust score for each contribution (rules) and overall trustworthiness of the CIDS node. As such, each CIDS node can conveniently infer the quality of the rules by looking at both scores.
    \item We develop a proof-of-concept implementation of our proposed architecture in a private Ethereum network hosted on a lab-scale testbed. We evaluate our architecture in terms of the evolution of trust scores, smart contract latency and Ethereum gas consumption. The results show that the our solution is feasible and performs within the expected benchmarks of the Ethereum platform.
\end{itemize}

The remainder of the paper is organised as follows. We present our proposed architecture with its underlying assumptions in Section~\ref{sec:proposed-model}. We outline the trust and reputation system in Section~\ref{sec:trust-and-reputation-module}, while the CIDS framework is discussed in Section~\ref{sec:cids-module}. We present the performance evaluation of our solution in Section~\ref{sec:performance-evaluation}. The related works are discussed in Section~\ref{sec:related-works} and we conclude our work in Section~\ref{sec:conclusion}.

\begin{figure*}
\centering
\includegraphics[width=0.9\textwidth]{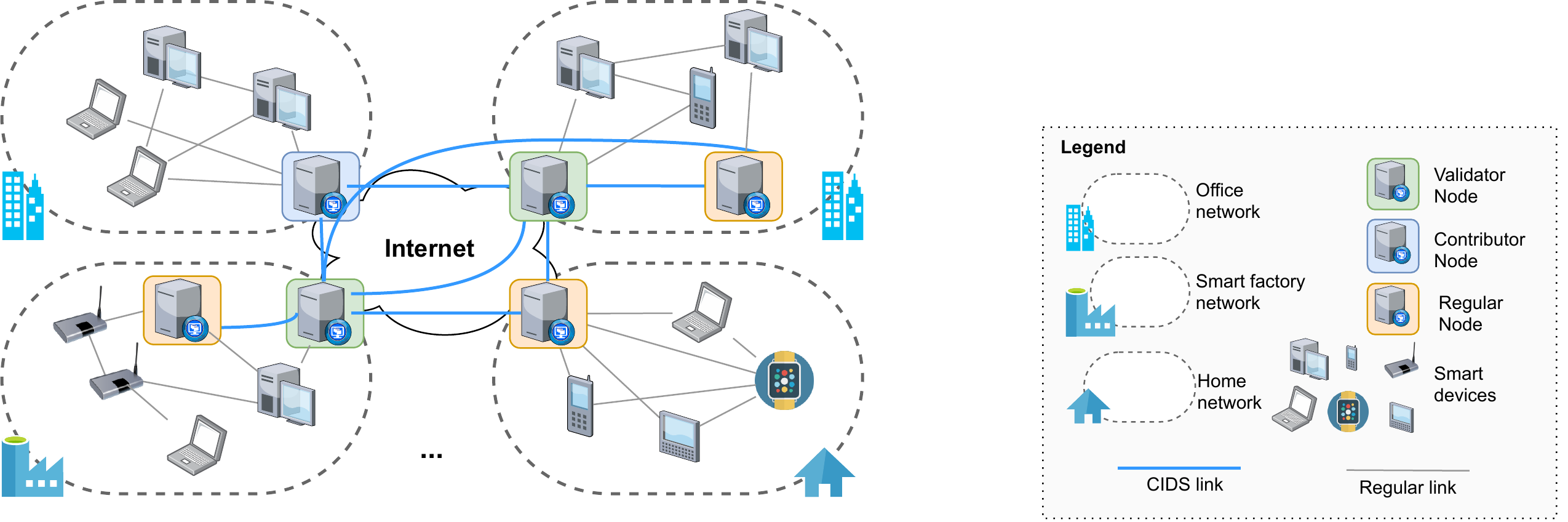}
\caption{An overview of the proposed decentralised CIDS.}
\label{fig:architecture-diagram}
\end{figure*}

\section{Proposed Architecture and Assumptions}
\label{sec:proposed-model}
In this section, we describe our decentralised CIDS architecture by elaborating the fundamental components and outlining the threat model. Lastly, we explain the assumptions in the present study.

\subsection{Architectural Overview}
\label{sub:system-model}
We present the overview of our proposed system in Figure~\ref{fig:architecture-diagram}. We design our architecture to span across multiple organisational networks, each of which typically comprises multiple smart devices with varying computational and storage capacity. In each organisation, we require at least a single CIDS node which acts as a signature-based IDS node that monitors the network for any attack occurrence. CIDS nodes communicate with other CIDS nodes through the TCP/IP protocol suite with an industry-standard encryption mechanism to ensure security, e.g., Transport Layer Security (TLS).

We model our system as $\mathbb{C}=(C, \mathbb{B}, \mathbb{I})$, where $C$ is a set of collaborating CIDS nodes that share intrusion related information, and $\mathbb{B}$ is the blockchain and $\mathbb{I}$ is the decentralised storage network. In our model, each CIDS node is connected to both blockchain $\mathbb{B}$ and decentralised storage network $\mathbb{I}$ for collaboration and storing shared detection rules, which contain a pattern of malicious network attacks, such as file hashes, malicious domains or particular byte sequences.
The blockchain network $\mathbb{B}$ is the main hub for collaboration, through which each CIDS node exchanges information via blockchain transactions. Each CIDS node collaboratively  builds a global trusted rules database $R_{db}$ that contains a collection of detection rules contributed by each CIDS node. Each CIDS node also keeps a local copy of an IDS database $R_{loc} \in R_{db}$. We summarise important notations used in this paper in Table~\ref{ta:notation-table}.

\begin{table}
\centering
\caption{Important notations and their description.}
\label{ta:notation-table}
\begin{tabular}{|c|l|}
\hline
\textit{\textbf{Notations}} & \textit{\textbf{Description}} \\
\hline \hline
$C$, $\mathbb{B}$, $\mathbb{I}$ & CIDS, blockchain and dec. storage network\\
\hline
$cv$, $cc$, $cr$ & a validator, contributor and regular node\\
\hline
$k_p$ and $k_s$ & the public and secret key\\
\hline
$r_{a,j}$ & an IDS rule from $cc_a$ \\
\hline
$S^{\, e}_{a,j}$ & validity score of $r_{a,j}$ from $cv_e$ \\
\hline
$t_{a,j}$ and $T^{\, m}_a$ & the trust score of $r_{a,j}$ and $cc_a$\\
\hline
$R_{db}$ and $R_{loc}$ & the global and local rules database \\
\hline
$SC_{trm}$ and $SC_{str}$ & the TRM and storage smart contracts \\
\hline
$E_{r_{a,j}}^v$ & new rule blockchain event\\
\hline
$E_{r_{a,j}}^o$ & new validated rule blockchain event\\
\hline
$\varphi^e_j$ & validation result from $cv_e$\\
\hline
$M_{r_{a,j}}$ & description of $r_{a,j}$ according to IDMEF~\cite{rfc4765}\\
\hline
$Z_{cc_a}$ & a zip archive of $M_{r_{a,j}}$ and $r_{a,j}$\\
\hline
$D_c$ & decision rule function \\
\hline
\end{tabular}
\end{table}

    \subsubsection{CIDS network}
    \label{subsub:cids-layer}
    All CIDS nodes form the CIDS network, which is connected to the other components of the system, e.g., blockchain network $\mathbb{B}$ and decentralised storage network $\mathbb{I}$. We assume that all CIDS nodes are equipped with sufficient resources to run a blockchain client. Each CIDS node holds a corresponding blockchain public-private key pair $\{k_p, k_s\}$ and is identifiable by the public key $k_p$. We define three types of IDS nodes, denoted $C=(C_V, C_C, C_R)$, where $C_V$, $C_C$ and $C_R$ correspond to validator, contributor and regular nodes, respectively.
    
    \textbf{Validator nodes:}
    Validator nodes, denoted $C_V=\{cv_1, cv_2, \ldots, cv_n\}$, are in charge of maintaining the CIDS network, including registration of new CIDS nodes and validation of contributed detection rules from other nodes by means of off-chain processes~\cite{Li2019}. Validator nodes validate detection rule $r_{a,j}$ submitted by other CIDS nodes via a consensus algorithm, which assigns a trust score to each rule. Validator nodes then append the validated rules to the trusted rules database $R_{db}$.

    \textbf{Contributor nodes:}
    We define a contributor node $\{cc_a | \forall cc \in C_C\}$ as a CIDS node that contributes a detection rule $r_{a,j}$ to the CIDS network $\mathbb{C}$. Each $cc_a$ has a trust score $T_a$ derived from the trustworthiness of its contributions.
    
    \textbf{Regular nodes:}
    We refer to the rest of the CIDS nodes in $C$ as regular nodes, denoted $C_R=\{cr_1, cr_2, \ldots, cr_k\}$. Regular nodes are only interested in using validated IDS rules in $R_{db}$ to update their local rules database $R_{loc}$ by subscribing to the system for a notification of newly validated $r_{a,j}$ in $R_{db}$. A $cr$ can conveniently examine the trust scores of both rule $r_{a,j}$ and node $cc$ to decide which rules to be included into their $R_{loc}$ based on a locally determined threshold. A regular node is passive and consequently does not get assigned a trust score.
    
    \subsubsection{Blockchain network}
    \label{subsub:blockchain-layer}
    We design our architecture to be blockchain agnostic, which supports any commodity blockchain platform with a prerequisite of supporting Turing-complete smart contract execution~\cite{solidity066}. In this work, we consider a single permissioned blockchain $\mathbb{B}$, wherein access is limited to certain known parties, which helps to build the first layer of defence. To avoid malicious actions, registration of new CIDS node is handled by $C_V$. In general, blockchain $\mathbb{B}$ is used to track the contributions of each $cc$ and to store the metadata of the contributed detection rules. Note that, we do not store the IDS rules on-chain but in the decentralised storage layer.
    
    We deploy two smart contracts onto blockchain $\mathbb{B}$. First, a Trust and Reputation Management (TRM) smart contract $SC_{trm}$ manages the contributions of all $cc$ by quantifying them via a transparent and verifiable TRM mechanism (see Section~\ref{sec:trust-and-reputation-module}). Second, a storage smart contract $SC_{str}$ maintains the hash and metadata of each $r_{a,j}$ to provide a connection between the blockchain and the decentralised storage network.
    
    \subsubsection{Decentralised storage network}
    \label{subsub:storage-layer}
    We employ IPFS, the InterPlanetary File System, as the decentralised storage network~\cite{benet2014ipfs}. We separate the data storage to prevent the blockchain size from becoming too large and thus ensuring scalability. The decentralised storage network is managed by $C_V$, which limits the access only to subscribed $cr$.

\subsection{Threat Model and Assumptions}
\label{sub:threat-model}
In our architecture, we assume that the adversaries are able to launch a poisoning attack aimed to inject misleading detection rules to the trusted database $R_{db}$. The adversaries can also compromise a maximum of $\ell$ validator nodes. However, we require that there are $n$ available validator nodes such that $n=3\ell+1$ to tolerate $\ell$ faulty validator node(s), i.e., $1/3$ fault tolerance as in PFBT~\cite{castro1999practical}.
The adversaries are also capable of performing trust and reputation attacks as follows:

\begin{itemize}
    \item \textit{Self-promoting attacks:} A malicious actor may try to increase its own reputation score by submitting and validating a detection rule by itself.
    \item \textit{Bad-mouthing attacks:} A malicious node may attempt to ruin the reputation of another node providing negative validation results regardless of the quality of the contributed model.
    \item \textit{Ballot-stuffing attacks:} A node may collude with other nodes to deliberately increase their reputation scores, for instance by submitting the same detection rules multiple times.
    \item \textit{Whitewashing or newcomer attacks:} A node attempts to rejoin the network using a new identity aiming to reset its previously recorded bad behaviour and obtain a fresh reputation score.
\end{itemize}

We assume that our system inherits the assumptions of a commodity blockchain platform, which include security against peer-to-peer and consensus attacks, such as Sybil, eclipse and 51\% attacks~\cite{8543246}. We assume that each validator holds a sufficient local detection rules database to validate the submitted detection rules and detect whether a submitted rule is malicious. All CIDS nodes in our architecture, regardless of their type, are bound to cryptographic primitives, which prevents manipulation and duplication of blockchain identities, i.e., public and private key pairs.

\section{Trust and Reputation Management}
\label{sec:trust-and-reputation-module}
We design the Trust and Reputation Management (TRM) to evaluate the trustworthiness of each contributor node $cc$ in the network $\mathbb{C}$ and also to protect the network from any corrupted or malicious detection rules. Each $cc$ will obtain a trust score after submitting a detection rule from which the score is calculated. In general, the score is increased when $cc$ contributes valid detection rules and decreased when submitting compromised rules. Intuitively, a high trust score indicates a trustworthy contributor node, while a low score indicates otherwise.

\begin{figure}[!t]
\centering
\includegraphics[width=0.43\textwidth]{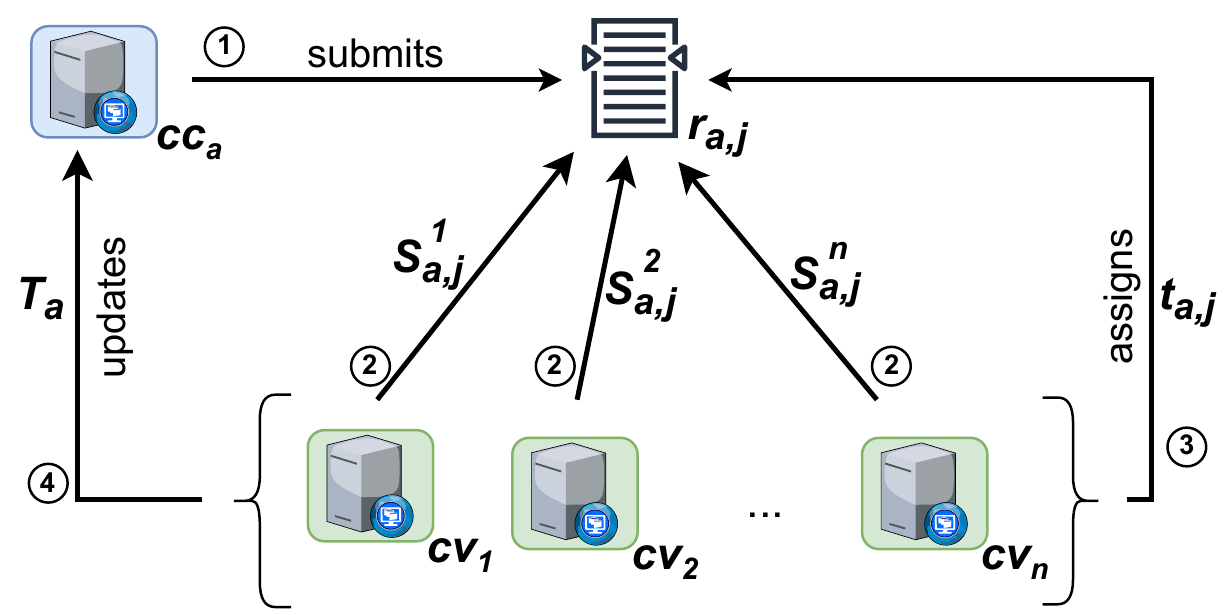}
\caption{The trust relationship model.}
\label{fig:trust-model}
\end{figure}

Figure~\ref{fig:trust-model} shows the trust relationship model in our proposed TRM. Suppose a contributor node $cc_a {\in} \, C_C$ submits a detection rule $r_{a,j}$. A validator node $cv_e$ validates $r_{a,j}$, and based on its accuracy, gives a score $S^{\, e}_{a,j} \in [0.5, 1]$ to indicate valid rules, or $S^{\, e}_{a,j} \in [0,0.5)$ to indicate invalid rules. As such, $S^{\, e}_{a,j}$ functions as a vote from $cv_e$ and consequently we have $n$ scores $\{S^1_{a,j}, S^2_{a,j}, \ldots, S^n_{a,j}\}$ for each $r_{a,j}$ as there are $n$ available validator nodes. We use the following formula to calculate the aggregated trust value of $r_{a,j}$, denoted by $t_{a,j}$:
\begin{equation}
\label{eq:rule-score}
    t_{a,j} = \frac{1}{n} \sum_{e=1}^{n} S^e_{a,j} \delta_e
\end{equation}
where
\begin{equation*}
    \delta_e = \begin{cases} 
                 \delta_{val}, & \text{if\ } S^{\, e}_{a,j} \geq 0.5 \\
                 \delta_{inv}, & \text{otherwise,}
             \end{cases}    
\end{equation*}
where $\delta_{val}$ and $\delta_{inv}$ are non-negative weights associated with a valid and invalid rule, which can be determined via a heuristic method. To favour trustworthy contributions over malicious ones, we set $\delta_{val} < \delta_{inv} \leq 1$ so that malicious contributions are assigned a higher weight. We propose a mechanism to collaboratively determine valid rules based on the scores from each validator node, which is elaborated in Section~\ref{sec:cids-module}. A valid $r_{a,j}$ is appended to $R_{db}$, while invalid $r_{a,j}$ would not be included.

Without loss of generality, let us assume that the first $q$  votes are valid and the rest are invalid, then we get
\begin{equation}
\label{eq:trustValue}
    t_{a,j} = \frac{1}{n} \left(\delta_{val}\sum_{e=1}^{q} S^e_{a,j} + \delta_{inv}\sum_{e=q+1}^{n} S^e_{a,j}\right)
\end{equation}
Since $S^{\, e}_{a,j} \in [0.5, 1]$ and $S^{\, e}_{a,j} \in [0,0.5)$ for valid and invalid rules respectively, we get the following bounds 
\begin{equation}
\label{eq:left-side}
    \frac{q \delta_{val}  }{2} \leq  \delta_{val}\sum_{e=1}^{q} S^e_{a,j} \leq q \delta_{val}
\end{equation}
and
\begin{equation}
\label{eq:right-side}
     0 \leq \delta_{inv}\sum_{e=q+1}^{n} S^e_{a,j} < \frac{\delta_{inv} (n-q)}{2}
\end{equation}
Then from~\eqref{eq:trustValue},~\eqref{eq:left-side}, and~\eqref{eq:right-side} we get the lower and upper bounds for $t_{a,j}$ as 
\begin{alignat}{2}
\label{eq:taj_bound}
& 0 \le t_{a,j} \le \delta_{val}, \quad && \text{for \,} \delta_{inv}<2\delta_{val} \nonumber\\
& 0 \le t_{a,j} < \delta_{inv}/2, \quad && \text{for \,} \delta_{inv}\geq2\delta_{val}
\end{alignat}


Subsequently, the trustworthiness score $T_a^{\, m}$ of $cc_a$ after contributing $m$ detection rules can be calculated based on the quality of each contributed rule:
\begin{equation}
\label{eq:trust}
    T_a^{\, m} = (1 - \gamma)\sum_{j = 1}^{m} \gamma^{(m - j)}t_{a,j}
\end{equation}
where $0 < \gamma \leq 1$ is the decaying constant to give more weights to recent contributions relative to older ones. To get the lower and upper bounds for $T_a^{\, m}$, let us assume that $t_{a,j}$ is constant. Then, from~\eqref{eq:trust} we get
\begin{align}
    T_a^{\, m} &= \frac{1 - \gamma^m}{1 - \gamma} (1 - \gamma) t_{a,j} \nonumber \\
    &= (1 - \gamma^m) t_{a,j}. \label{sub-eq:trust}
\end{align}
From~\eqref{eq:taj_bound} and~\eqref{sub-eq:trust}, we get the lower and upper bounds of $T_{a}^{\, m}$ as
\begin{alignat}{2}
\label{eq:tam_limit}
& 0 \le T_a^{\,m} \le (1 - \gamma^m)\delta_{val}, \quad && \text{for \,} \delta_{inv}<2\delta_{val} \nonumber\\
& 0 \le T_a^{\,m} < (1 - \gamma^m)\delta_{inv}/2, \quad && \text{for \,} \delta_{inv}\geq2\delta_{val}
\end{alignat}

We implement the trust calculation for both $t_{a,j}$ and $T_a^{\,m}$ in the $SC_{trm}$ smart contract, using which each $cv$ can cooperate to calculate the score and collaboratively determine valid rules. In addition, the $SC_{trm}$ smart contract also holds the computed trust scores. As the calculation and storage are offloaded to blockchain $\mathbb{B}$, regular nodes $cr$ can conveniently query blockchain $\mathbb{B}$ and examine $t_{a,j}$ and $T_a^{\,m}$ scores to infer the quality of the detection rules, thus reducing the computation loads. We elaborate on how the TRM mechanism is used in practice in Section~\ref{sec:cids-module}.

\section{Collaborative Intrusion Detection System}
\label{sec:cids-module}
\begin{figure}[t]
\centering
\includegraphics[width=0.47\textwidth]{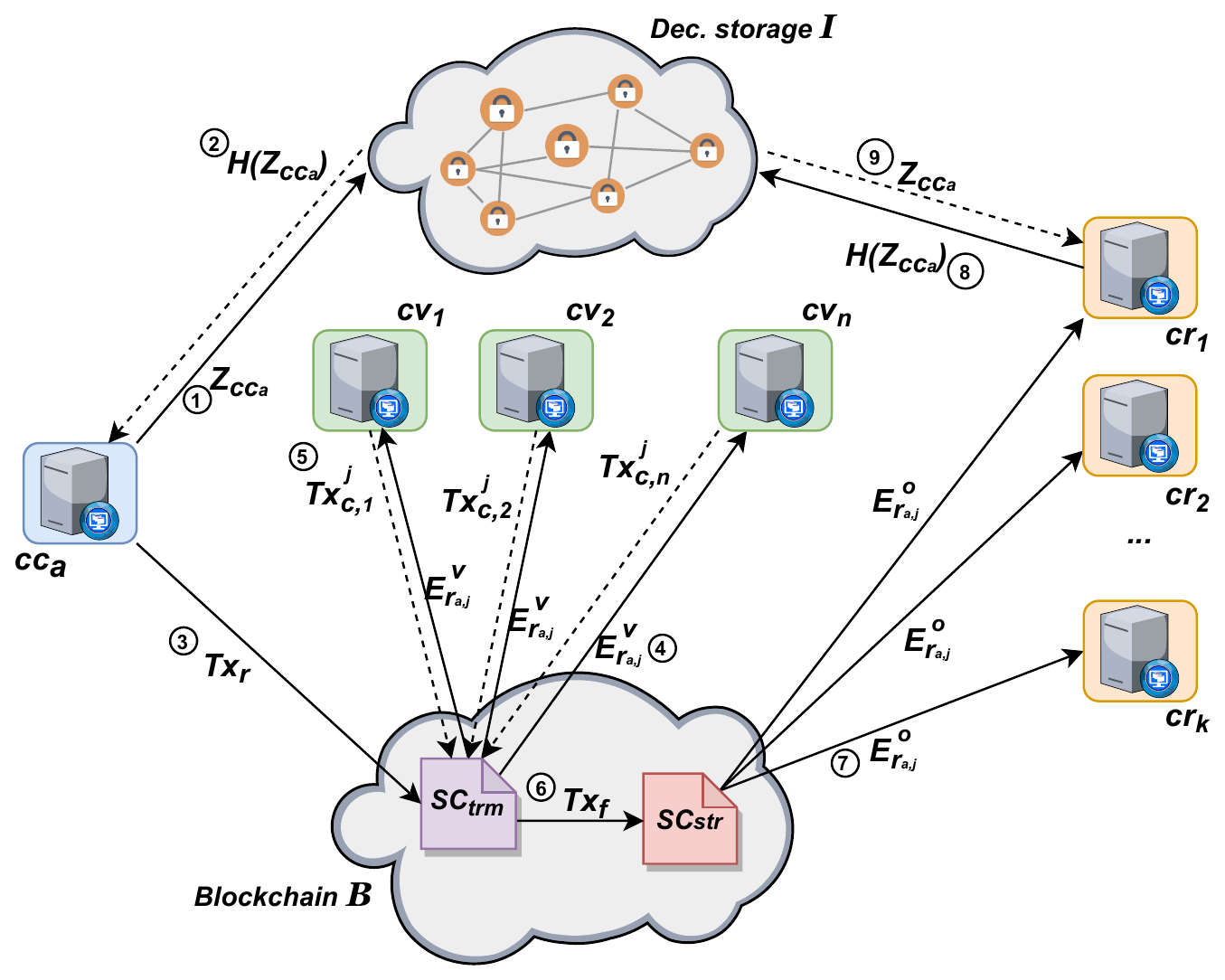}
\caption{The workflow of the proposed CIDS platform.}
\label{fig:sequence-diagram}
\end{figure}

In this section, we describe the mechanism of our proposed CIDS framework. We present the overview of the underlying processes as a sequence diagram in Figure~\ref{fig:sequence-diagram}, which displays the process from when $cc_a$ submits $r_{a,j}$ until the rule is validated and stored in the rules database. Refer to Table~\ref{ta:notation-table} for a summary of the notations used in this paper.

We assume there is a set of $n$ online validator nodes $C_V=\{cv_1, cv_2, \ldots, cv_n\}$ to initialise the network $\mathbb{C}$ and to deploy the smart contracts $SC_{trm}$ and $SC_{str}$. As we support less than $1/3$ of the online $cv$ to be byzantine nodes, we require at least $n=4$ online $cv$ for PBFT fault tolerance~\cite{castro1999practical}. Note that the CIDS network $\mathbb{C}$ is a private network, thereby all CIDS nodes should be known in advance, although they are not necessarily trusted. We presume that each new CIDS node is able to generate a public-private key pair $\{k_p, k_s\}$. A new CIDS node can join $\mathbb{C}$ as a regular node by sending a request $Req_{cr}$ over a secure channel to any $cv$:
\begin{equation}
    Req_{cr} = \left<k_p, attr_{cr}, timestamp, Sign_{cr}\right>
\end{equation}
where $k_p$ is the node's public key, $attr_{cr}$ is an attribute which also includes its IP address and a unique identifier, while $Sign_{cr}$ is the signature of the message. Once the request is approved, the corresponding validator node will give a response containing the details for IPFS connection (e.g., the IP address and the hash address of the bootstrap node) and the addresses of both $SC_{trm}$ and $SC_{str}$, using which the nodes can access the network rules database $R_{db}$ and obtain updates when a new detection rule has been added to $R_{db}$.

$SC_{trm}$ facilitates collaborative rule validation to determine whether a new detection rule should be included to $R_{db}$ based on the votes from each $cv$. To submit a new rule $r_{a,j}$, a $cc_a$ node is required to invoke a function on $SC_{trm}$ which then triggers a blockchain event to notify all regulator nodes $\forall cr \in C_R$ about the newly submitted rule. Note that this smart contract-based consensus mechanism is not intended to replace the built-in consensus algorithm as it serves a different purpose and runs on top of the built-in consensus algorithm of the underlying blockchain platform. Although the validator nodes are relatively trusted nodes in $\mathbb{C}$, we incorporate our rule validation consensus mechanism to mitigate if some $cv$'s are compromised. Next, we explain our proposed rule validation consensus mechanism.

Let $cc_a$ be a contributor node that is about to contribute a new detection rule $r_{a,j}$ to $R_{db}$. Firstly, $cc_a$ needs to add a zip file $Z_{cc_a} = \left< r_{a,j}, M_{r_{a,j}} \right>$ to the decentralised storage network $\mathbb{I}$, where $M_{r_{a,j}}$ is the description of $r_{a,j}$. To provide compliance and interoperability, we follow Intrusion Detection Message Exchange Format (IDMEF)~\cite{rfc4765} as the format of $M_{r_{a,j}}$. Subsequently, $cc_a$ obtains the hash address of $Z_{cc_a}$, denoted $H(Z_{cc_a})$, required for file retrieval. $cc_a$ is then required to invoke transaction $Tx_r$ to $SC_{trm}$:
\begin{equation}
\label{eq:tx_r}
    Tx_r = [\, H(Z_{cc_a}) \parallel timestamp \parallel Sig_{cc_a} \,]
\end{equation}
where $Sig_{cc_a}$ is the signature on $hash(H(Z_{cc_a}) {\parallel} timestamp)$ using the signing key $k_{s_{cc}}$ for authentication. Subsequently, $SC_{trm}$ triggers a blockchain event $E_{r_{a,j}}^v$ to notify all validator nodes that a new rule $r_{a,j}$ has been added to the queue for validation.

Using the hash address $H(Z_{cc_a})$, each validator node $cv$ retrieves and extracts $Z_{cc_a}$ from storage $\mathbb{I}$ for off-chain validation~\cite{Li2019}. Here, each $cv$ compares $r_{a,j}$ against its local database to confirm whether $r_{a,j}$ performs as per $M_{r_{a,j}}$. The validator also inspects if $r_{a,j}$ has been previously submitted either by the same $cc_a$ or another node to avoid ballot-stuffing attack. Depending on the validation results, a validator node $cv_e$ may either approve or reject $r_{a,j}$, along with a score $S^{\, e}_{a,j}$ that indicates the quality of $r_{a,j}$. To submit the vote about the validity of $r_{a,j}$, all validator nodes submit $Tx_{c,e}^j$ to $SC_{trm}$:
\begin{equation}
\label{eq:tx_c}
    Tx_{c,e}^j = [\, \varphi^e_j \parallel S^{\, e}_{a,j} \parallel timestamp \parallel Sig_{cv_e} \,]
\end{equation}
where $\varphi^e_j \in \{1, -1\}$ is the validation result to indicate a valid ($1$) or invalid rule ($-1$) and $Sig_{cv_e}$ is the signature on $hash(\varphi^e_j {\parallel} S^{\, e}_{a,j} {\parallel} timestamp)$ using signing key $k_{s_{cv}}$, which is used for authentication. We adapt a weighted majority rule~\cite{Leonardos2020} to make a decision on the validity of $r_{a,j}$. We define a decision rule $D_c : \mathbf{x}_j \rightarrow \{-1, 1\}$ which receives $\mathbf{x}_j = (Tx_{c,1}^j, Tx_{c,2}^j, \ldots, Tx_{c,n}^j)$ as an input and outputs a decision $\{1, -1\}$, where $1$ and $-1$ indicate valid and invalid rules, respectively. We define $D_c$ as follows:
\begin{equation}
\label{eq:weighted-decision-rule}
    D_c(\mathbf{x}_j) := \begin{cases} 
                 \phantom{-}1, & \text{if\ }  \nicefrac{1}{n}\sum_{e=1}^{n} S^e_{a,j} \varphi^e_j  \geq q \text{\ ,}\\
                 -1, & \text{otherwise}
             \end{cases}    
\end{equation}
where $q \in (0, 1]$ is the threshold of a valid rule, which can be determined via a heuristic method.

$SC_{trm}$ proceeds with the decision making process, once all votes from all $cv$ have been received, as described in Algorithm~\ref{alg:cids-algorithm}. A valid $r_{a,j}$ would be appended to $R_{db}$, while an invalid $r_{a,j}$ would be ignored. $SC_{trm}$ calculates the trust score $t_{a,j}$ and $T^{\, m}_a$ as defined in~\eqref{eq:rule-score} and~\eqref{eq:trust}, respectively. Subsequently, $SC_{trm}$ invokes $Tx_f$ to $SC_{str}$:
\begin{equation}
\label{eq:tx_f}
    Tx_f = [\, H(Z_{cc_a}) \parallel t_{a,j} \parallel timestamp \,]
\end{equation}
to include the newly approved $r_{a,j}$ to the rules database $R_{db}$. To notify all regular nodes about the new rule, $SC_{str}$ triggers a blockchain event $E_{r_{a,j}}^o$. A regular node $cr$ can now retrieve $Z_{cc_a}$ from $\mathbb{I}$ using the address $H(Z_{cc_a})$. After inspecting the score $t_{a,j}$ and metadata $M_{r_{a,j}}$, $cr$ may opt to include $r_{a,j}$ into its local IDS rules database $R_{loc}$ for better detection of new attacks.

Unlike typical consensus mechanisms where each device sends messages to each other (broadcast message), in our proposed mechanism, the participating nodes are only required to send a message to the smart contract, which acts as an aggregator. Thus, our proposed mechanism would work on any blockchain instantiation that supports smart contracts for on-chain logic execution.

\begin{algorithm}[t]
\caption{Consensus mechanism for rule validation.}
\label{alg:cids-algorithm}
\begin{algorithmic}[1]
\renewcommand{\algorithmicensure}{\textbf{Output:}}
\Require $Tx_c$, $n$, $q$ and $C_R$
\Ensure  $Tx_f$ or $\emptyset$
\State $r\_count \leftarrow getState()$ \Comment{from the blockchain}
\If{$r\_count$ == $(n-1)$} \Comment{all validation received}
    \State $calculate \ t_{a,j}$ \Comment{as in~\eqref{eq:rule-score}}
    \State $update \ T^{\, m}_a$ \Comment{as in~\eqref{eq:trust}}
    \State $calculate \ D_c(\mathbf{x}_j)$ \Comment{as in~\eqref{eq:weighted-decision-rule}}
    \If {$D_c(\mathbf{x}_j) == 1$}
        \State \Return $Tx_f$ \Comment{$r_{a,j}$ is added to $R_{db}$}
    \Else
        \State \Return $\emptyset$ \Comment{$r_{a,j}$ is rejected}
    \EndIf
\Else
    \State $saveToState(r\_count$++$)$ \Comment{write to blockchain}
    \State \Return $\emptyset$
\EndIf

\end{algorithmic}
\end{algorithm}



\section{Performance Evaluation}
\label{sec:performance-evaluation}
This section presents the performance evaluation of the proposed CIDS architecture based on our proof of concept (POC) implementation. We first describe the details of the POC that was implemented on a lab-scale private Ethereum network. Then, we present the experimental results to show the feasibility of our solution, which include trust evaluation for honest and malicious nodes and the blockchain performance with regards to latency and gas consumption. 

\begin{figure*}
    \centering
    \begin{tabularx}{\linewidth}{XXX}
        \includegraphics[width=0.32\textwidth]{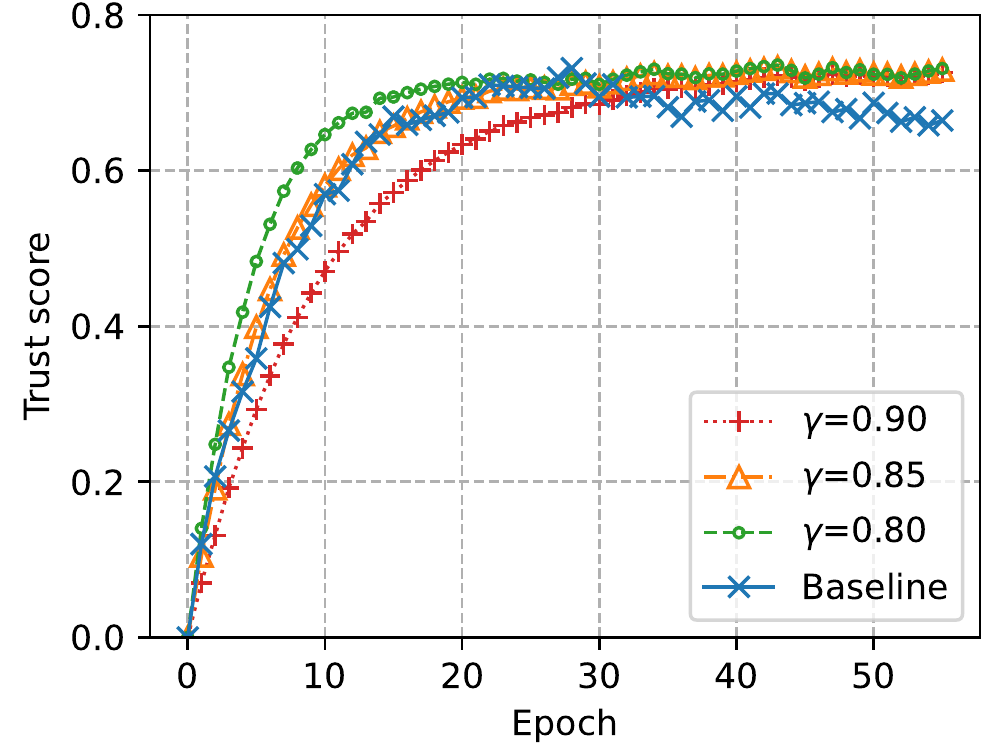}
        \caption{Convergence of $T_{a}^{\,m}$ for an honest $cc$ with varying $\gamma$, $\delta_{val} = 0.85$ and baseline~\cite{Kolokotronis2019}.}
        \label{fig:trust_evolution}
        &
        \includegraphics[width=0.32\textwidth]{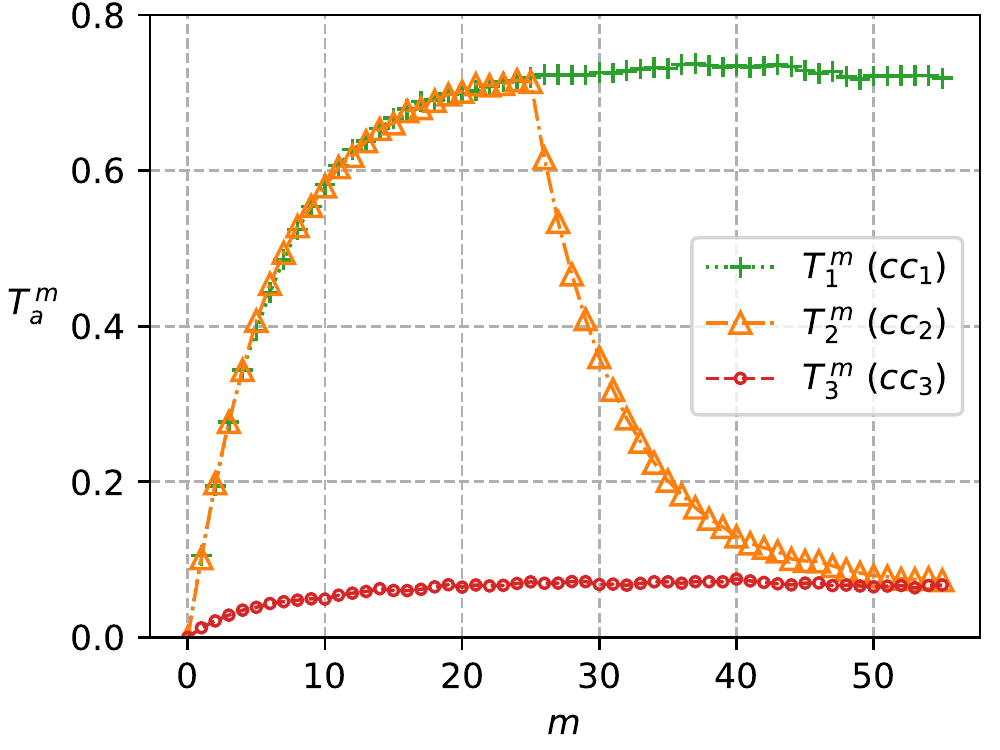}
        \caption{Trust score comparison for an honest and malicious $cc$ ($\delta_{val} = 0.85$, $\delta_{inv} = 0.9$, $\gamma = 0.85$).}
        \label{fig:honest_malicious}
        &
        \includegraphics[width=0.32\textwidth]{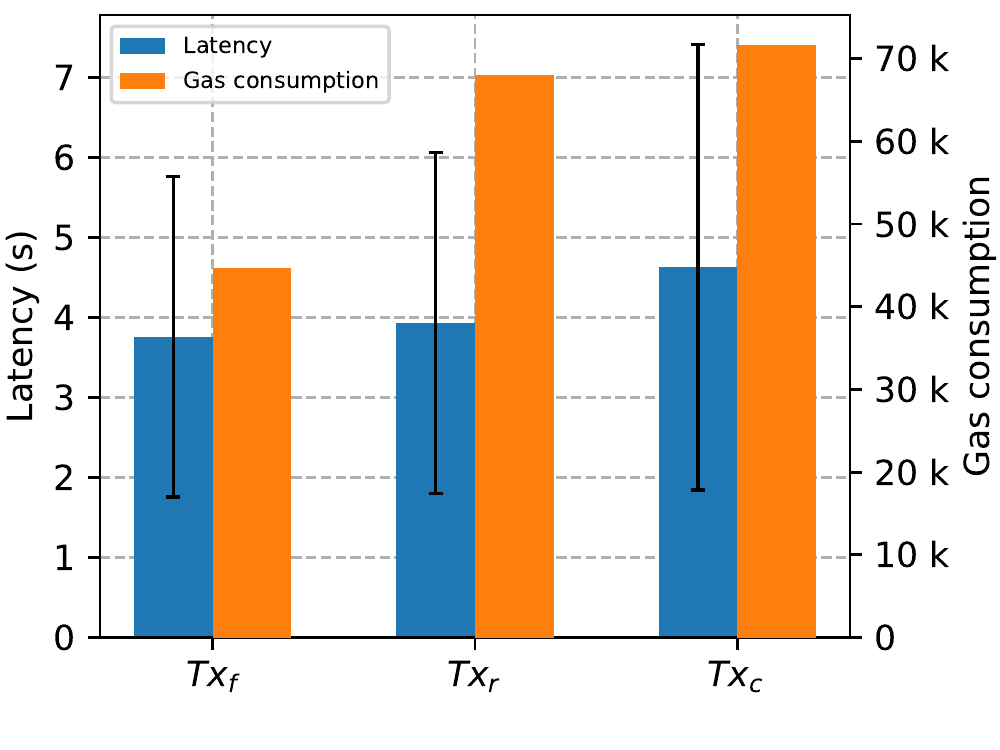}
        \caption{Comparison of the latency and gas consumption for rule confirmation, validation and submission.}
        \label{fig:gas_consumption}
    \end{tabularx}
\end{figure*}

\begin{figure}
\centering
\includegraphics[width=0.43\textwidth]{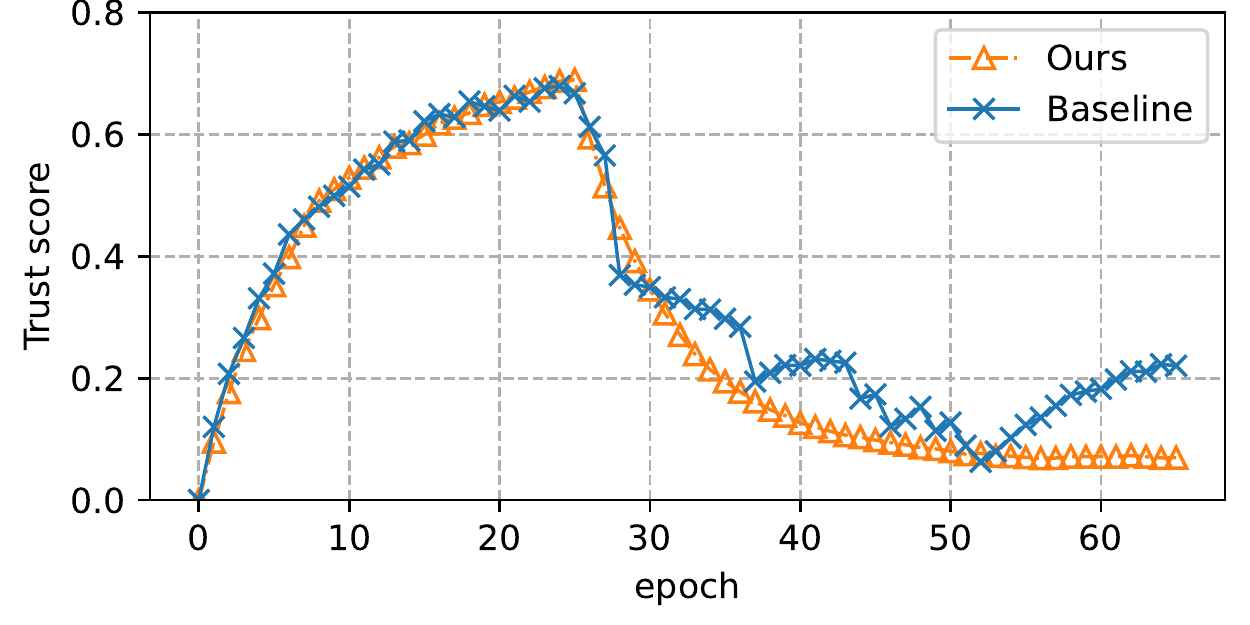}
\caption{Quantitative comparison of trust evolution with baseline~\cite{Kolokotronis2019}.}
\label{fig:baseline-comparison}
\end{figure}

\subsection{Implementation Details}
\label{sub:implementation-details}
We opted for a private Ethereum blockchain as our POC platform, as Ethereum natively supports smart contracts written in a Turing complete programming language, Solidity~\cite{solidity066}. Ethereum utilises a decentralised environment named Ethereum Virtual Machine (EVM), as a trusted and secure platform for decentralised computation~\cite{wood2014ethereum}.
However, we note that our proposed CIDS architecture can also be implemented in any blockchain platform that offers support for smart contracts, for instance Hyperledger Fabric and Sawtooth.

We built a lab-scale testbed which comprises 15 nodes of Raspberry Pi with different specifications (7 Raspberry Pi 3 and 8 Raspberry Pi 4) and a Lenovo ThinkCentre mini PC (8GB RAM, 2.9 GHz quad-core Intel Core i5 CPU) as the platform to run our experiments. We selected Raspberry Pi computers to imitate the different smart devices with varying capacity, while we used the Lenovo ThinkCentre as a more powerful node. To orchestrate the CIDS nodes, we utilise Docker containers\footnote{\url{https://docs.docker.com/engine/}} running on the testbed using which we simulated a total of $16$ nodes: $4$ validator, $4$ contributor, and $8$ regular nodes. We built a private Ethereum network with a single miner and a private IPFS network for the decentralised storage. We utilise an open-source IDS, Snort\footnote{\url{https://www.snort.org/}}, as the signature-based IDS platform, and geth as the Ethereum client. We wrote a Python v3.8.10 and a bash script to control and simulate interactions between CIDS nodes with web3py v5.11 and py-solc v3.2.0 libraries as the middleware for Ethereum connection and smart contracts compilation, respectively. We implemented the smart contracts, i.e., $SC_{trm}$ and $SC_{str}$ in Solidity v0.6.6~\cite{solidity066}, which offers built-in libraries for computation and verification of hashes efficiently.


\subsection{Evaluation Results}
\label{sub:evaluation-results}
The following subsections present the experimental results of our POC implementation. We compare our results against challenge-based trust mechanism as the baseline~\cite{Kolokotronis2019}, where a CIDS node sends challenge messages to another CIDS node and waits for the responses (e.g., valid, invalid, unsure), using which the trustworthiness level is derived. To mimic similar characteristics with our proposal (e.g., trust score range), we set the values of the baseline's parameters as: forgetting factor $\lambda = 0.85$, severity level $\phi = 2$ and initial trust score $\tau = 0$.

\subsubsection{Convergence of the trust score}
\label{subsub:trust-convergence}
We evaluate the convergence of the trust score $T_{a}^{\,m}$ by varying the decaying constant $\gamma$ and plot the results in Figure~\ref{fig:trust_evolution}. This experiment simulates an honest $cc$ which constantly contributes $m=55$ valid detection rules. As we vary the value of $\gamma$ from $0.9$ to $0.8$, different weights are allocated for the latest and the previous $t_{a,j}$ values in calculating the final $T_{a}^{\,m}$. However, as all contributions are valid, $T_{a}^{\,m}$ converges to a similar level as illustrated in Figure~\ref{fig:trust_evolution}, regardless of the $\gamma$ value, which confirms the theoretical upper bound of $T_{a}^{\,m}$, described in~\eqref{eq:tam_limit}. A subtle difference is that lower values of $\gamma$ would result in a gradual growth of $T_{a}^{\,m}$, thereby $\gamma$ can be tuned according to the practical settings. Figure~\ref{fig:trust_evolution} also shows a minor fluctuation of $T_{a}^{\,m}$, as $T_{a}^{\,m}$ depends on $t_{a.i}$ values, which are weighted averages of different scores $S^e_{a,j}$ from each validator $cv_e$. Relative to the baseline, our trust evolution exhibits similar trend of positive interaction with less fluctuations.

\subsubsection{Trust evolution of honest and malicious nodes}
\label{subsub:honest-malicious}
We simulate three contributor nodes ($cc_1$, $cc_2$ and $cc_3$) to examine the evolution of $T_{a}^{\,m}$ for different node behaviours over $m=55$ contributions. First, $cc_1$ is an honest contributor node that consistently contributes trustworthy rules. Second, $cc_2$ initially submits trustworthy rules up to $m=25$ and turns malicious by contributing false rules afterwards. Third, $cc_3$ is a compromised node that constantly submits poor detection rules for the entire simulation. We apply $\delta_{val} = 0.85$, $\delta_{inv} = 0.9$, $\gamma = 0.85$ as the parameters and plot the experimental results in Figure~\ref{fig:honest_malicious}. While all $T_{a}^{\,m}$ initially begins at $0$, $T_{1}^{\,m}$ and $T_{2}^{\,m}$ continue to grow gradually in a similar rate and start to plateau approximately at $T_{a}^{\,m} = 0.75$. On the other hand, $T_{3}^{\,m}$ saturates at approximately $T_{a}^{\,m} = 0.1$, as the validators are assigning very low $S^e_{a,j}$ scores (non-zero). At $m>25$, $T_{2}^{\,m}$ declines significantly and saturates at a similar score of $T_{3}^{\,m}$ for the rest of the experiment. We note that the scores are in line with the theoretical bounds of $T_{a}^{\,m}$, as described in~\eqref{eq:taj_bound} and~\eqref{eq:tam_limit}.

We plot the evolution of $T_{a}^{\,m}$ against the baseline in Figure~\ref{fig:baseline-comparison}. Similar to Figure~\ref{fig:honest_malicious}, node $cc_2$ initially acts honestly by sending valid responses until $25$-th interaction, resulting in high trust scores, and turns malicious by sending invalid and \textit{unsure} responses for epoch time $> 25$. Our trust evolution exhibits stable and gradual growth and decline when the node acts honestly and maliciously, respectively. However, the baseline suffers from undesired fluctuations when the node acts maliciously, i.e., epoch time $> 25$, as \textit{unsure} responses perturb the slope of the curve, cf. Section 3.B of~\cite{Kolokotronis2019}.


\subsubsection{Latency and gas consumption}
\label{subsub:latency-gas-consumption}
Ethereum requires a fee for executing blockchain transactions, with Gas as the unit of measurement.
The Gas is calculated based on the required EVM opcodes during the execution of smart contract's functions~\cite{solidity066}.
To examine the feasibility of our blockchain implementation, we measure the gas consumption and execution latency for the following smart contract functions: 1) rule submission ($Tx_r$), 2) rule validation ($Tx_c$), and 3) rule confirmation ($Tx_f$). We repeated the experiments 30 times and plot the results in Figure~\ref{fig:gas_consumption}. The average execution latencies are relatively similar for all transactions, which fall within the range of $3$ and $5$ seconds. There is a relatively high variance of the latencies, as there is no assurance of when the miner processes and appends the transaction to the blockchain. $Tx_f$ consumes the least Gas, while $Tx_c$ consumes the most Gas among these three transactions, due to the different amount of required EVM opcodes (cf.~\eqref{eq:tx_f}, \eqref{eq:tx_r}, \eqref{eq:tx_c}).

\section{Related Works}
\label{sec:related-works}
There have been several proposals in the literature about blockchain incorporation in CIDS. Li et al. proposed a framework called CBSigIDS which aims to avoid insider attacks, where malicious nodes can generate untruthful signatures to contaminate the network~\cite{Li2019}. 
Here, blockchain is incorporated to provide a mechanism for sharing detection signatures between different IDS nodes in a verifiable manner.
CBSigIDS also implements trust computation to evaluate the reputation levels of the IDS nodes by means of a challenge-based trust mechanism, wherein the CIDS nodes are required to send challenge messages to known acquaintances to assess the trustworthiness level in detecting known attacks. While the authors incorporated blockchain, CBSigIDS does not utilise smart contracts to take full advantage of blockchain.

In~\cite{Kolokotronis2019}, blockchain is incorporated in a CIDS solution to store and disseminate calculated trust scores with the underlying evidence that justifies the trust calculation. The solution mainly aims to enhance the overall security by recording misbehaving CIDS nodes. Similar to~\cite{Li2019}, the solution employs a challenge-based trust evaluation. However, the solution does not explore incentives and penalties as a reward mechanism and no performance evaluation has been undertaken.

Blockchain has also been proposed for building a CIDS framework in Software Defined Network (SDN) environment where SDN controllers also act as CIDS nodes~\cite{Fan2020}. The framework employs permissioned blockchain to guard against adversaries that attempt to manipulate detection signatures by sharing untruthful detection signatures to the participating SDN controllers. The authors proposed digital certificates to establish trust between SDN controllers by designing a scheme called Certificate-Chain (C-Chain) which is essentially a blockchain-based distributed Public Key Infrastructure (PKI). However, the framework does not quantitatively evaluate the SDN controllers' trustworthiness. In addition, the framework utilises IPFS as a storage medium for detection signature files. However, the framework does not impose penalties for fraudulent manipulations.

Alkadi et al. proposed a CIDS platform for cloud based IoT networks, wherein blockchain facilitates immutable data exchange between several cloud services~\cite{Alkadi2020}. In contrast to signature-based methods, the platform combines blockchain with a Deep Learning (DL) technique to provide a secure CIDS with smart contracts in cloud based IoT networks. The platform utilises a consortium blockchain with a Trusted Execution Environment (TEE) for securely logging the transactions between multiple cloud providers, while a Bidirectional Long Short-Term Memory (BiLSTM) DL algorithm is trained to detect anomalies in the network. However, this DL platform does not incorporate trust management for evaluating cloud providers (CIDS nodes) and does not explore incentive mechanisms for the collaborations.

Another proposal that uses DL for blockchain-based CIDS is presented in~\cite{Liang2021}. The authors designed a collaboration framework where blockchain is employed to consistently train and test detection models for anomaly-based CIDS. The proposed blockchain-based CIDS aims to achieve a lifetime learning framework, which is able to gradually build a secure and co-maintained database of labelled training set for classification. The framework adopts Growing Hierarchical Self-Organising Map with probabilistic relabelling (GHSOM-pr) as the off-chain classifier, which can adapt to the dynamic nature of the co-maintained database. While the framework introduces Data Coins (DCoins) as the incentives for collaboration, it does not incorporate smart contracts and include mechanisms for trustworthiness evaluation.

It is worth noting that even though various techniques have been proposed to secure CIDS networks, some issues remain unsolved. For instance, current trust management mechanisms for CIDS employ a challenge-based method that was initially designed for Host-based IDS (HIDS)~\cite{Fung2011}. Challenge-based trust management requires each node to calculate and store trust scores on the device itself, which could be seen as redundant. While the scheme may work well for medium sized networks, challenge-based techniques would be impractical and raise scalability issues when the number of CIDS nodes are relatively large.

\section{Conclusion}
\label{sec:conclusion}
In this paper, we proposed a trust framework to build a trustworthy CIDS architecture that continuously evaluates the trustworthiness of the CIDS nodes with regard to the detection rules contributed by each IDS node. We used signature-based CIDS as an illustrative example, while our architecture is blockchain agnostic and could be implemented on any blockchain platform that supports smart contracts. We built a lab-scale testbed to evaluate our proposed architecture in a private Ethereum network. Our experimental results showed the feasibility of our concept and that the performance falls within the expected benchmarks of the Ethereum platform. For future work, we aim to evaluate the economic model and game theory behind the fees and incentives mechanism and conduct more extensive evaluation of our architecture.

\section*{Acknowledgements}
\noindent The authors acknowledge the support of the Commonwealth of Australia and Cyber Security Cooperative Research Centre for this work.

\bibliographystyle{IEEEtran}
\bibliography{./bibliography/mybib}

\begin{thebibliography}{10}
\providecommand{\url}[1]{#1}
\csname url@samestyle\endcsname
\providecommand{\newblock}{\relax}
\providecommand{\bibinfo}[2]{#2}
\providecommand{\BIBentrySTDinterwordspacing}{\spaceskip=0pt\relax}
\providecommand{\BIBentryALTinterwordstretchfactor}{4}
\providecommand{\BIBentryALTinterwordspacing}{\spaceskip=\fontdimen2\font plus
\BIBentryALTinterwordstretchfactor\fontdimen3\font minus
  \fontdimen4\font\relax}
\providecommand{\BIBforeignlanguage}[2]{{%
\expandafter\ifx\csname l@#1\endcsname\relax
\typeout{** WARNING: IEEEtran.bst: No hyphenation pattern has been}%
\typeout{** loaded for the language `#1'. Using the pattern for}%
\typeout{** the default language instead.}%
\else
\language=\csname l@#1\endcsname
\fi
#2}}
\providecommand{\BIBdecl}{\relax}
\BIBdecl

\bibitem{Anthi2019}
E.~Anthi, L.~Williams, M.~Slowinska, G.~Theodorakopoulos, and P.~Burnap, ``{A
  Supervised Intrusion Detection System for Smart Home IoT Devices},''
  \emph{IEEE IoT Journal}, vol.~6, no.~5, pp. 9042--9053, 2019.

\bibitem{Fung2011}
C.~J. Fung, J.~Zhang, I.~Aib, and R.~Boutaba, ``{Dirichlet-based trust
  management for effective collaborative intrusion detection networks},''
  \emph{IEEE TNSM}, vol.~8, no.~2, pp. 79--91, 2011.

\bibitem{Li2019}
W.~Li, S.~Tug, W.~Meng, and Y.~Wang, ``{Designing collaborative blockchained
  signature-based intrusion detection in IoT environments},'' \emph{Future
  Generation Computer Systems}, vol.~96, 2019.

\bibitem{Liang2021}
J.~Liang and M.~Ma, ``Co-maintained database based on blockchain for idss: A
  lifetime learning framework,'' \emph{IEEE TNSM}, vol.~18, no.~2, pp.
  1629--1645, 2021.

\bibitem{Golomb2018}
T.~Golomb, Y.~Mirsky, and Y.~Elovici, ``{CIoTA: Collaborative IoT Anomaly
  Detection via Blockchain},'' in \emph{Workshop on Decentralized IoT Security
  and Standards (DISS) 2018}, 2018.

\bibitem{9097612}
G.~D. {Putra}, V.~{Dedeoglu}, S.~S. {Kanhere}, and R.~{Jurdak}, ``Towards
  scalable and trustworthy decentralized collaborative intrusion detection
  system for iot,'' in \emph{5th IEEE/ACM IoTDI}, 2020, pp. 256--257.

\bibitem{rfc4765}
H.~Debar, D.~Curry, and B.~Feinstein, ``{The Intrusion Detection Message
  Exchange Format (IDMEF)},'' Internet Requests for Comments, {RFC Editor},
  {RFC} 4765, March 2007.

\bibitem{solidity066}
Solidity, ``{Solidity - Solidity 0.6.6 documentation},''
  \url{https://solidity.readthedocs.io/en/v0.6.6/}, 2020, [Online; accessed
  4-June-2020].

\bibitem{benet2014ipfs}
J.~Benet, ``Ipfs - content addressed, versioned, p2p file system,'' 2014.

\bibitem{castro1999practical}
M.~Castro, B.~Liskov \emph{et~al.}, ``Practical byzantine fault tolerance,'' in
  \emph{OSDI}, vol.~99, no. 1999, 1999, pp. 173--186.

\bibitem{8543246}
M.~A. {Ferrag}, M.~{Derdour}, M.~{Mukherjee}, A.~{Derhab}, L.~{Maglaras}, and
  H.~{Janicke}, ``Blockchain technologies for the internet of things: Research
  issues and challenges,'' \emph{IEEE IoT Journal}, vol.~6, no.~2, pp.
  2188--2204, 2019.

\bibitem{Leonardos2020}
S.~Leonardos, D.~Reijsbergen, and G.~Piliouras, ``Weighted voting on the
  blockchain: Improving consensus in proof of stake protocols,''
  \emph{International Journal of Network Management}, vol.~30, no.~5, 2020.

\bibitem{Kolokotronis2019}
N.~Kolokotronis, S.~Brotsis, G.~Germanos, C.~Vassilakis, and S.~Shiaeles, ``{On
  blockchain architectures for trust-based collaborative intrusion
  detection},'' \emph{IEEE SERVICES 2019}, pp. 21--28, 2019.

\bibitem{wood2014ethereum}
G.~Wood \emph{et~al.}, ``Ethereum: A secure decentralised generalised
  transaction ledger,'' \emph{Ethereum project yellow paper}, vol. 151, no.
  2014, pp. 1--32, 2014.

\bibitem{Fan2020}
W.~Fan, Y.~Park, S.~Kumar, P.~Ganta, X.~Zhou, and S.-y. Chang,
  ``{Blockchain-enabled Collaborative Intrusion Detection in Software Defined
  Networks},'' in \emph{19th IEEE TrustCom}, 2020, pp. 967--974.

\bibitem{Alkadi2020}
O.~Alkadi, N.~Moustafa, B.~Turnbull, and K.-K.~R. Choo, ``{A Deep Blockchain
  Framework-enabled Collaborative Intrusion Detection for Protecting IoT and
  Cloud Networks},'' \emph{IEEE IoT-J}, vol. 4662, 2020.

\end{thebibliography}

\end{document}